\documentclass[titlepage,12pt]{article}
\usepackage[dvips]{graphicx}

\begin{document}

\title{Klein-Gordon particles in mixed vector-scalar inversely linear potentials}
\date{}
\author{Antonio S. de Castro \\
\\
UNESP - Campus de Guaratinguet\'{a}\\
Departamento de F\'{\i}sica e Qu\'{\i}mica\\
Caixa Postal 205\\
12516-410 Guaratinguet\'{a} SP - Brasil\\
\\
E-mail address: castro@feg.unesp.br (A.S. de Castro)}
\date{}
\maketitle

\begin{abstract}
The problem of a spinless particle subject to a general mixing of vector and
scalar inversely linear potentials in a two-dimensional world is analyzed.
Exact bounded solutions are found in closed form by imposing boundary
conditions on the eigenfunctions which ensure that the effective Hamiltonian
is Hermitian for all the points of the space. The nonrelativistic limit of
our results adds a new support to the conclusion that even-parity solutions
to the nonrelativistic one-dimensional hydrogen atom do not exist.
\end{abstract}

The problem of a particle subject to an inversely linear potential in one
spatial dimension ($\sim |x|^{-1}$), known as the one-dimensional hydrogen
atom, has received considerable attention in the literature (for a rather
comprehensive list of references, see \cite{xia}). This problem presents
some conundrums regarding the parities of the bound-state solutions and the
most perplexing is that one regarding the ground state. Loudon \cite{lou}
claims that the nonrelativistic Schr\"{o}dinger equation provides a
ground-state solution with infinite eigenenergy and a related eigenfunction
given by a delta function centered about the origin. This problem was also
analyzed with the Klein-Gordon equation for a Lorentz vector coupling and
there it was revealed a finite eigenenergy and an exponentially decreasing
eigenfunction \cite{spe}. By using the technique of continuous
dimensionality the problem was approached with the Schr\"{o}dinger,
Klein-Gordon and Dirac equations \cite{mos}. The conclusion in this last
work reinforces the claim of Loudon. Furthermore, the author of Ref. \cite
{mos} concludes that the Klein-Gordon equation, with the interacting
potential considered as a time component of a vector, provides unacceptable
solutions while the Dirac equation has no bounded solutions at all. On the
other hand, in a more recent work \cite{xia} the authors use connection
conditions for the eigenfunctions and their first derivatives across the
singularity of the potential, and conclude that only the odd-parity
solutions of the Schr\"{o}dinger equation survive. The problem was also
sketched for a Lorentz scalar potential in the Dirac equation \cite{ho}, but
the analysis is incomplete. In a recent work \cite{asc6} it was shown that
the problem of a fermion under the influence of a general scalar potential
for nonzero eigenenergies can be mapped into a Sturm-Liouville problem.
Next, the key conditions for the existence of bound-state solutions were
settled for power-law potentials, and the possible zero-mode solutions were
shown to conform with the ultrarelativistic limit of the theory. In
addition, the solution for an inversely linear potential was obtained in
closed form. The effective potential resulting from the mapping has the form
of the Kratzer potential \cite{kra}. It is noticeable that this problem has
an infinite number of acceptable bounded solutions, nevertheless it has no
nonrelativistic limit for small quantum numbers. It was also shown that in
the regime of strong coupling additional zero-energy solutions can be
obtained as a limit case of nonzero-energy solutions. The ideas of
supersymmetry had already been used to explore the two-dimensional Dirac
equation with a scalar potential \cite{coo}-\cite{nog}, nevertheless the
power-law potential has been excluded of such discussions.

The Coulomb potential of a point electric charge in a 1+1 dimension,
considered as the time component of a Lorentz vector, is linear ($\sim
\left| x\right| $) and so it provides a constant electric field always
pointing to, or from, the point charge. This problem is related to the
confinement of fermions in the Schwinger and in the massive Schwinger models
\cite{col1}-\cite{col2}, and in the Thirring-Schwinger model \cite{fro}. It
is frustrating that, due to the tunneling effect (Klein\'{}s paradox), there
are no bound states for this kind of potential regardless of the strength of
the potential \cite{cap}-\cite{gal}. The linear potential, considered as a
Lorentz scalar, is also related to the quarkonium model in one-plus-one
dimensions \cite{hoo}-\cite{kog}. Recently it was incorrectly concluded that
even in this case there is solely one bound state \cite{bha}. Later, the
proper solutions for this last problem were found \cite{cas1}-\cite{hil}.
However, it is well known from the quarkonium phenomenology in the real 3+1
dimensional world that the best fit for meson spectroscopy is found for a
convenient mixture of vector and scalar potentials put by hand in the
equations (see, e.g., \cite{luc}). The same can be said about the treatment
of the nuclear phenomena describing the influence of the nuclear medium on
the nucleons \cite{ser}-\cite{nosso}. The mixed vector-scalar potential has
also been analyzed in 1+1 dimensions for a linear potential \cite{cas2} as
well as for a general potential which goes to infinity as $|x|\rightarrow
\infty $ \cite{ntd}. In both of those last references it has been concluded
that there is confinement if the scalar coupling is of sufficient intensity
compared to the vector coupling.

Motived by the success found in Ref. \cite{asc6}, the problem of a fermion
in the background of an inversely linear potential by considering a
convenient mixing of vector and scalar Lorentz structures was re-examined
\cite{asc7}. The problem was mapped into an exactly solvable Sturm-Liouville
problem of a Schr\"{o}dinger-like equation with an effective Kratzer
potential. The case of a pure scalar potential with their isolated
zero-energy solutions, already analyzed \cite{asc6}, was obtained as a
particular case. In the present paper the same problem is analyzed for a
spinless particle. These new results favour the conclusion that even-parity
solutions to the nonrelativistic one-dimensional hydrogen atom do not exist.

In the presence of vector and scalar potentials the 1+1 dimensional
time-independent Klein-Gordon equation for a spinless particle of rest mass $%
m$ reads

\begin{equation}
-\hbar ^{2}c^{2}\frac{d^{2}\psi }{dx^{2}}+\left( mc^{2}+V_{s}\right)
^{2}\psi =\left( E-V_{v}\right) ^{2}\psi  \label{1b}
\end{equation}

\noindent where $E$ is the energy of the particle, $c$ is the velocity of
light and $\hbar $ is the Planck constant. The vector and scalar potentials
are given by $V_{v}$ and $V_{s}$, respectively. The subscripts for the terms
of potential denote their properties under a Lorentz transformation: $v$ for
the time component of the 2-vector potential and $s$ for the scalar term. It
is worth to note that the Klein-Gordon equation is covariant under $%
x\rightarrow -x$ if $V_{v}(x)$ and $V_{s}(x)$ remain the same.

In the nonrelativistic approximation (potential energies small compared to $%
mc^{2}$ and $E\simeq mc^{2}$) Eq. (\ref{1b}) becomes

\begin{equation}
\left( -\frac{\hbar ^{2}}{2m}\frac{d^{2}}{dx^{2}}+V_{v}+V_{s}\right) \psi
=\left( E-mc^{2}\right) \psi  \label{1e}
\end{equation}

\noindent Eq. (\ref{1e}) shows that $\psi $ obeys the Schr\"{o}dinger
equation with binding energy equal to $E-mc^{2}$, and without distinguishing
the contributions of vector and scalar potentials.

It is remarkable that the Klein-Gordon equation with a scalar potential, or
a vector potential contaminated with some scalar coupling, is not invariant
under $V\rightarrow V+const.$, this is so because only the vector potential
couples to the positive-energies in the same way it couples to the
negative-ones, whereas the scalar potential couples to the mass of the
particle. Therefore, if there is any scalar coupling the absolute values of
the energy will have physical significance and the freedom to choose a
zero-energy will be lost. It is well known that a confining potential in the
nonrelativistic approach is not confining in the relativistic approach when
it is considered as a Lorentz vector. It is surprising that relativistic
confining potentials may result in nonconfinement in the nonrelativistic
approach. This last phenomenon is a consequence of the fact that vector and
scalar potentials couple differently in the Klein-Gordon equation whereas
there is no such distinction among them in the Schr\"{o}dinger equation.
This observation permit us to conclude that even a ``repulsive'' potential
can be a confining potential. The case $V_{v}=-V_{s}$ presents bounded
solutions in the relativistic approach, although it reduces to the
free-particle problem in the nonrelativistic limit. The attractive vector
potential for a particle is, of course, repulsive for its corresponding
antiparticle, and vice versa. However, the attractive (repulsive) scalar
potential for particles is also attractive (repulsive) for antiparticles.
For $V_{v}=V_{s}$ and an attractive vector potential for particles, the
scalar potential is counterbalanced by the vector potential for
antiparticles as long as the scalar potential is attractive and the vector
potential is repulsive. As a consequence there is no bounded solution for
antiparticles. For $V_{v}=0$ and a pure scalar attractive potential, one
finds energy levels for particles and antiparticles arranged symmetrically
about $E=0$. For $V_{v}=-V_{s}$ and a repulsive vector potential for
particles, the scalar and the vector potentials are attractive for
antiparticles but their effects are counterbalanced for particles. Thus,
recurring to this simple standpoint one can anticipate in the mind that
there is no bound-state solution for particles in this last case of mixing.

Now let us focus our attention on scalar and vector potentials in the form
\begin{equation}
V_{s}=-\frac{\hbar cq_{s}}{|x|},\quad V_{v}=-\frac{\hbar cq_{v}}{|x|}
\label{12}
\end{equation}
\noindent where the coupling constants, $q_{s}$ and $q_{v}$, are
dimensionless real parameters. In this case Eq. (\ref{1b}) becomes

\begin{equation}
H_{eff}\psi =-\frac{\hbar ^{2}}{2m}\,\psi ^{\prime \prime }+V_{eff}\,\psi
=E_{eff}\,\psi  \label{12a}
\end{equation}
where
\begin{equation}
E_{eff}=\frac{E^{2}-m^{2}c^{4}}{2mc^{2}},\quad V_{eff}=-\frac{\hbar cq_{eff}%
}{|x|}+\frac{A}{x^{2}}  \label{12b}
\end{equation}
\noindent and
\begin{equation}
q_{eff}=\,\frac{mc^{2}q_{s}+Eq_{v}}{mc^{2}},\quad A=\frac{\hbar ^{2}}{2m}%
\,\left( q_{s}^{2}-q_{v}^{2}\right)  \label{14}
\end{equation}

\noindent Therefore, one has to search for bounded solutions of a particle
in an effective Kratzer-like potential. The Klein-Gordon eigenvalues are
obtained by inserting the effective eigenvalues in (\ref{12b}).

Before proceeding, it is useful to make some qualitative arguments regarding
the Kratzer-like potential and its possible solutions. The effective
Kratzer-like potential is unable to bind particles on the condition that
either $q_{eff}<0$ and $A\geq 0$, or $q_{eff}=0$ and $A>0$, because the
effective potential is repulsive everywhere. For $q_{eff}<0$ and $A<0$,
there appears a double-barrier potential structure with singularity given by
$-1/x^{2}$. In such a case the effective potential could be a binding
potential even with $E_{eff}>0$. In all the remaining cases the effective
potential could bind the particle only with $E_{eff}<0$, corresponding to
Klein-Gordon eigenvalues in the range $-mc^{2}<E<+mc^{2}$. For $q_{eff}>0$
and $A>0$ there is a double-well potential structure, and for $q_{eff}>0$
and $A\leq 0$ as well as for $q_{eff}=0$ and $A<0$, there appears a
singularity at the origin given by $-1/x^{2}$ ($-1/|x|$ if $A=0$). It is
worthwhile to note at this point that the singularity $-1/x^{2}$ will never
expose the particle to collapse to the center \cite{lan} on the condition
that $A$ is never less than the critical value $A_{c}=-\hbar ^{2}/(8m)$,
i.e., $q_{s}^{2}\geq q_{v}^{2}-1/4$. This last observation allows us to get
rid of the incongruous possibility of bounded solutions for $q_{eff}<0$ and $%
A<0$ mentioned above, and for $q_{eff}=0$ and $A<0$ yet. Therefore, one can
foresee that only for $q_{eff}>0$, i.e., $q_{s}>-q_{v}E/(mc^{2})$, can the
potential hold bound-state solutions, and that the eigenenergies in the
range $|E|>mc^{2}$ correspond to the continuum.

The Schr\"{o}dinger equation with the Kratzer-like potential is an exactly
solvable problem and its solution, for an attractive inversely linear term
plus a repulsive inverse-square term in the potential, can be found on
textbooks \cite{lan}-\cite{flu}. Since we need solutions involving repulsive
as well as attractive terms in the potential, the calculation including this
generalization is presented. The spectrum will be uniquely determined by
ensuring the Hermiticity of the effective Hamiltonian given by (\ref{12a}).
This means that we will demand normalizable and orthogonal eigenfuntions. As
a bonus, the appropriate boundary conditions on the Klein-Gordon wave
functions will be proclaimed.

Defining the quantities $z$ and $B$,

\negthinspace
\begin{equation}
z=\frac{2}{\hbar }\sqrt{-2E_{eff}}\;|x|,\quad B=q_{eff}\,c\,\sqrt{-\frac{1}{%
2E_{eff}}}  \label{15}
\end{equation}
\noindent \noindent and using (\ref{12a})-(\ref{12b}) one obtains the
equation

\begin{equation}
\,\psi ^{\prime \prime }+\left( -\frac{1}{4}+\frac{B}{z}-\frac{2A}{\hbar
^{2}z^{2}}\right) \psi =0  \label{16}
\end{equation}

\noindent Now the prime denotes differentiation with respect to $z$. The
normalizable asymptotic form of the solution as $z\rightarrow \infty $ is $%
e^{-z/2}$. As $z\rightarrow 0$, when the term $1/z^{2}$ dominates, the
solution behaves as $z^{s}$, where $s$ is a solution of the algebraic
equation

\begin{equation}
s(s-1)-\frac{2A}{\hbar ^{2}}=0  \label{17}
\end{equation}
viz.

\begin{equation}
s=\frac{1}{2}\left( 1\pm \sqrt{1+\frac{8A}{\hbar ^{2}}}\right) =\frac{1}{2}%
\pm \sqrt{q_{s}^{2}-q_{v}^{2}+\frac{1}{4}}  \label{18}
\end{equation}

\noindent Notice that  $A\geq A_{c}$. The boundary conditions on the
eigenfunctions comes into existence by demanding that the effective
Hamiltonian given (\ref{12a}) is Hermitian, viz.

\begin{equation}
\int_{0}^{\infty }dx\;\psi _{n}^{*}\left( H_{eff}\psi _{n^{^{\prime
}}}\right) =\int_{0}^{\infty }dx\;\left( H_{eff}\psi _{n}\right) ^{*}\psi
_{n^{^{\prime }}}  \label{22-1}
\end{equation}

\noindent In passing, note that a necessary consequence of Eq. (\ref{22-1})
is that the eigenfunctions corresponding to distinct effective eigenvalues
are orthogonal. Recalling that $\psi _{n}(\infty )=0$, it can be shown that (%
\ref{22-1}) is equivalent to

\begin{equation}
\lim_{x\rightarrow 0}\left( \psi _{n}^{*}\frac{d\psi _{n^{^{\prime }}}}{dx}-%
\frac{d\psi _{n}^{*}}{dx}\psi _{n^{^{\prime }}}\right) =0  \label{22-2}
\end{equation}

\noindent There results that the allowed values for $s$ are restricted to $%
s\geq 1/2$ and
\begin{equation}
q_{s}^{2}\geq q_{v}^{2}-\frac{1}{4}  \label{22-3}
\end{equation}
The constraint over $s$ implies that only the positive sign in (\ref{18})
must be considered. The solution for all $z$ can be expressed as $\psi $ $%
(z)=z^{s}e^{-z/2}w(z)$, where $w$ is solution of Kummer\'{}s equation \cite
{abr}

\begin{equation}
zw^{\prime \prime }+(b-z)w^{\prime }-aw=0  \label{19}
\end{equation}

\noindent with

\begin{equation}
a=s-B,\quad b=2s  \label{20}
\end{equation}

\noindent Then $w$ is expressed as $M(a,b,z)$ and in order to furnish
normalizable $\psi $, the confluent hypergeometric function must be a
polynomial. This demands that $a=-n$, where $n$ is a nonnegative integer in
such a way that $M(a,b,z)$ is proportional to the associated Laguerre
polynomial $L_{n}^{b-1}(z)$, a polynomial of degree $n$. This requirement,
combined with the first equation of (\ref{20}), also implies into quantized
Klein-Gordon eigenvalues:

\begin{equation}
E=mc^{2}\,\frac{-\frac{q_{s}q_{v}}{\left( s+n-1\right) ^{2}}\pm \sqrt{1-%
\frac{q_{s}^{2}-q_{v}^{2}}{\left( s+n-1\right) ^{2}}}}{1+\frac{q_{v}^{2}}{%
\left( s+n-1\right) ^{2}}},\qquad n=1,2,3,\ldots  \label{21}
\end{equation}

\noindent The Klein-Gordon eigenfunctions on the half-line are given by

\begin{equation}
\psi (z)=Nz^{s}e_{\;}^{-z/2}\;L_{n-1}^{2s-1}\left( z\right)  \label{22}
\end{equation}

\noindent where $N$ is a constant related to the normalization.
Since $B$ is a positive number,
\begin{equation}
q_{s}>-q_{v}\,\frac{E}{mc^{2}}  \label{22a}
\end{equation}
\noindent as advertized by the preceding qualitative arguments.

Eqs. (\ref{22-3}) and (\ref{22a}) can be used to achieve the constraints on
the coupling constants as well as on the allowed signs of $E$. For instance,
there are only positive (negative) energy solutions when a $q_{v}=+q_{s}$ ($%
q_{v}=-q_{s}$) and $0<q_{s}<1$. Likewise, there are only positive (negative)
energy solutions if $q_{v}$ is positive (negative) and $q_{s}\leq 0$. In all
the other circumstances the spectra may acquiesce both signs of
eigenenergies. Anyway, $E\rightarrow -E$ and $\psi $ is invariant as $%
q_{v}\rightarrow -q_{v}$.

Now that we know the solution of the problem on the half-line, we will start
to analyze some illustrative particular cases.

\textbf{1)} $q_{v}=0$. In the case of a pure scalar potential Eq. (\ref{22a}%
) demands that $q_{s}>0$. The energy levels are given by
\begin{equation}
E=\pm mc^{2}\,\sqrt{1-\left( \frac{q_{s}}{s+n-1}\right) ^{2}}  \label{23d}
\end{equation}

\noindent so that the energy levels for particles and antiparticles are
symmetric about $E=0$. The nonrelativistic limit of the theory, a regime of
weak coupling ($q_{s}\ll 1$), furnishes $E-mc^{2}\approx
-mc^{2}q_{s}^{2}/(2n^{2})$. On the other hand, in the regime of strong
coupling, i.e., for $q_{s}\gg 1$, one has $E\approx \pm mc^{2}\left(
n-1\right) /q_{s}$ and as the coupling becomes extremely strong the lowest
effective eigenvalues end up close to zero. One sees clearly that the
eigenvalues for a zero-energy solution can be obtained only as a limit case
of a nonzero-energy solution.

\textbf{2)} $q_{v}=q_{s}$. The coupling constants are restricted to positive
values, and the energy levels given by

\begin{equation}
E=mc^{2}\,\frac{n^{2}-q_{s}^{2}}{n^{2}+q_{s}^{2}}  \label{23c}
\end{equation}

\noindent are pushed down from the upper continuum so that these energy
levels correspond to bound states of particles. In this case there are no
energy levels for antiparticles. All the Klein-Gordon eigenvalues are
positive if $q_{s}<1$, and some negative eigenvalues arise if $q_{s}>1$. One
has $E-mc^{2}\approx -2mc^{2}q_{s}^{2}/n^{2}$ as long as $q_{s}\ll 1$.

\textbf{3)} $q_{v}=-q_{s}$. Eq. (\ref{22a}) demands that $q_{s}>0$. Only the
energy levels emerging from the lower continuum, the energy levels for
antiparticles, survive:
\begin{equation}
E=-mc^{2}\,\frac{n^{2}-q_{s}^{2}}{n^{2}+q_{s}^{2}}  \label{23e}
\end{equation}

\noindent Note that $E\approx mc^{2}$ only in the strong-coupling regime.

\textbf{4)} $q_{s}=0$. In the case of a pure vector potential Eqs. (\ref
{22-3}) and (\ref{22a}) demand that $|q_{v}|\leq 1/2$ and $\varepsilon
(E)=\varepsilon (q_{v})$, where $\varepsilon $ stands for the sign function.
It follows that

\begin{equation}
E=\varepsilon (q_{v})\,\frac{mc^{2}}{\sqrt{1+\left( \frac{q_{v}}{s+n-1}%
\right) ^{2}}}  \label{24}
\end{equation}

\noindent In this circumstance, the energy spectrum consists of energy
levels either for particles ($q_{v}>0$) or for antiparticles ($q_{v}<0$).
The nonrelativistic limit of the theory, a regime of weak coupling ($%
0<q_{v}\ll 1/2$), furnishes $E-mc^{2}\approx -mc^{2}q_{v}^{2}/(2n^{2})$.

There are no bounded solutions for particles in mixture $q_{v}=-q_{s}$ and
the nonrelativistic limit is not viable, as expected. For all the other
particular cases one sees that the regime of weak coupling runs in the
appropriate nonrelativistic limit: the energy levels for particles are given
by the nonrelativistic Coulomb potential. As a matter of fact, this
nonrelativistic limit is always feasible in the regime of weak coupling
provided $q_{s}>-q_{v}$. In all the circumstances, there is no atmosphere
for the spontaneous production of particle-antiparticle pairs. No matter the
signs of the potentials or how strong they are, the particle and
antiparticle levels neither meet nor dive into the continuum. Thus there is
no room for the production of particle-antiparticle pairs. This all means
that Klein\'{}s paradox never comes to the scenario.

\begin{figure}[th]
\begin{center}
\includegraphics[width=9cm, angle=270]{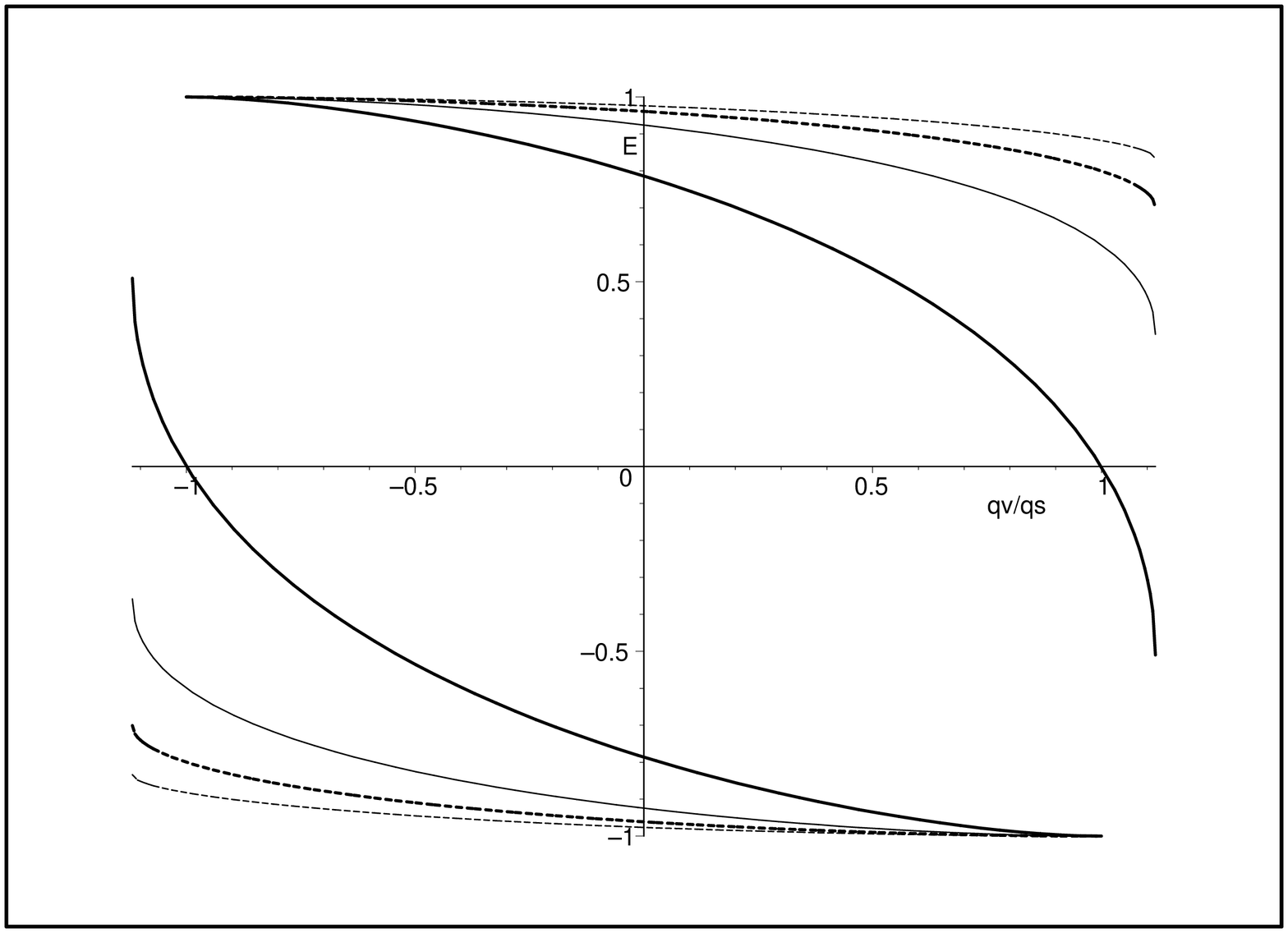}
\end{center}
\par
\vspace*{-0.1cm}
\caption{Klein-Gordon eigenvalues for the four lowest energy levels as a
function of $q_{v} /q_{s}$. The full thick line stands for $n=1$, the full
thin line for $n=2$, the heavy dashed line for $n=3$ and the light dashed
line for $n=4$ ($m=c=q_{s}=1$). }
\label{Fig1}
\end{figure}

\begin{figure}[th]
\begin{center}
\includegraphics[width=9cm, angle=270]{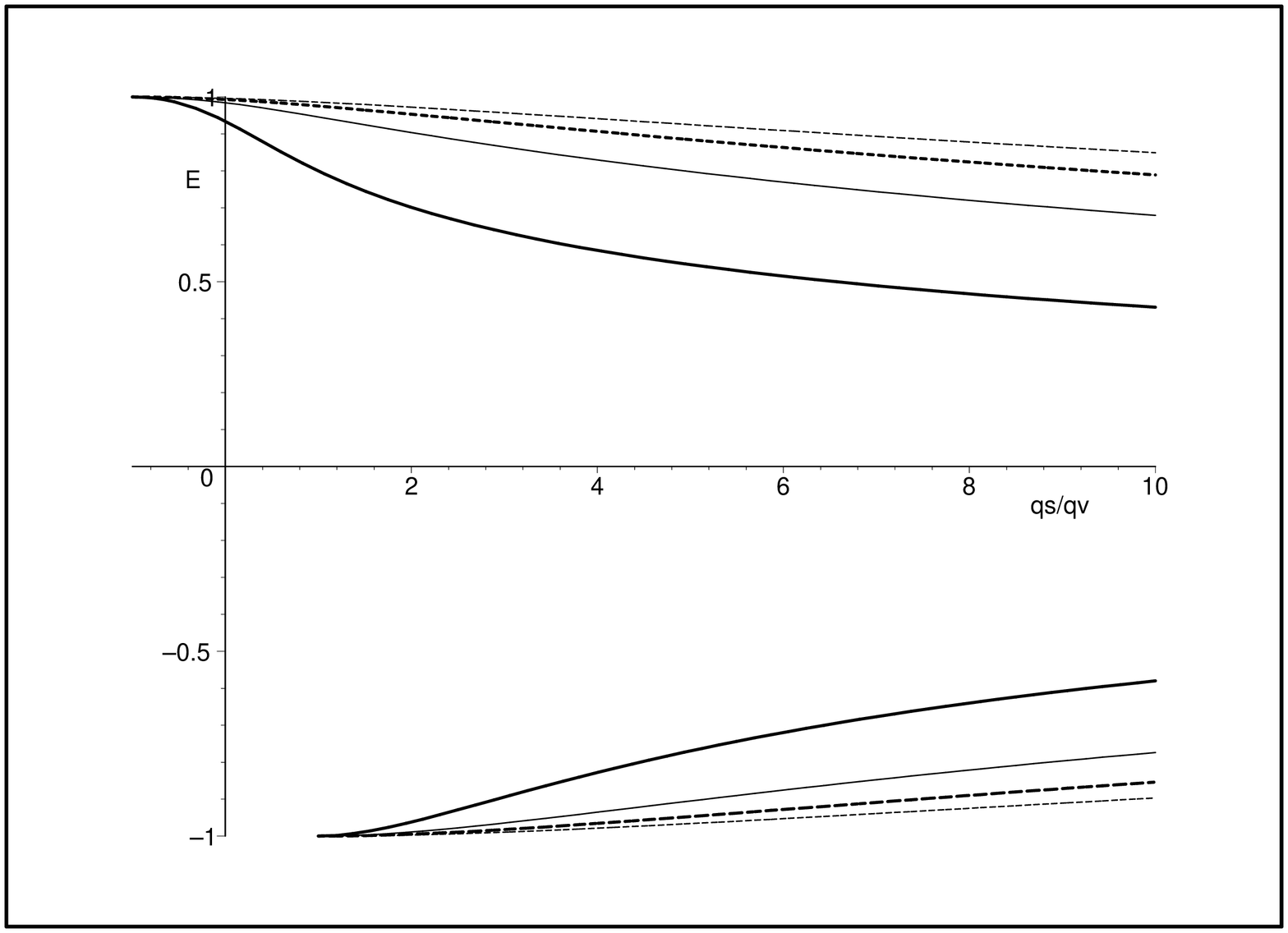}
\end{center}
\par
\vspace*{-0.1cm}
\caption{Klein-Gordon eigenvalues for the four lowest energy levels as a
function of $q_{s} /q_{v}$. The full thick line stands for $n=1$, the full
thin line for $n=2$, the heavy dashed line for $n=3$ and the light dashed
line for $n=4$ ($m=c=q_{v}=1/3$). }
\label{Fig2}
\end{figure}

\begin{figure}[th]
\begin{center}
\includegraphics[width=9cm, angle=270]{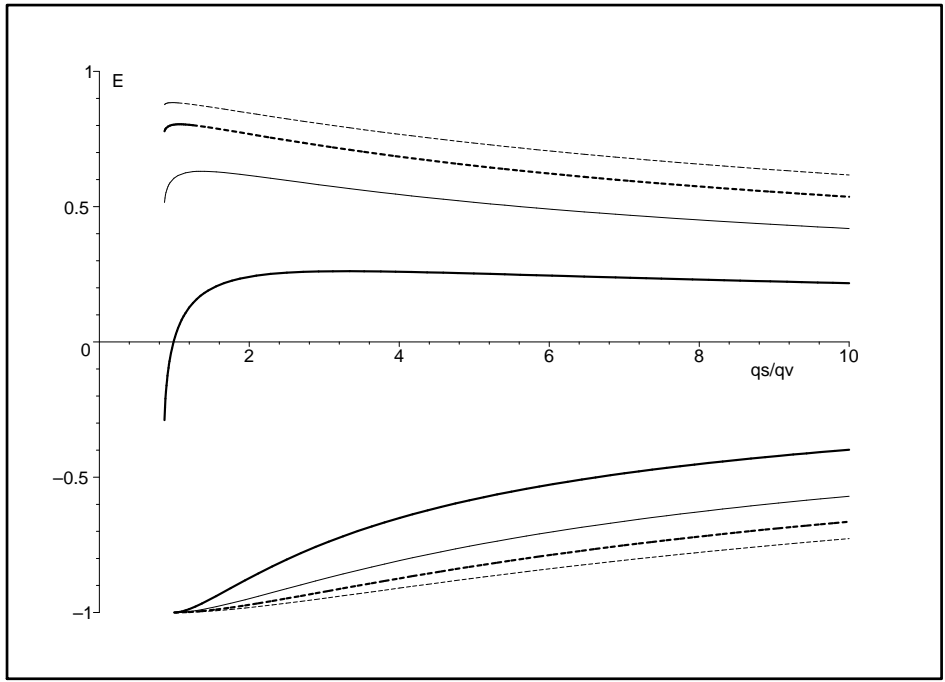}
\end{center}
\par
\vspace*{-0.1cm}
\caption{The same as in Fig. 2, for $q_{v}=1$.}
\label{Fig3}
\end{figure}

The Klein-Gordon eigenenergies are plotted in Fig. \ref{Fig1} for the four
lowest bound states as a function of $q_{v}/q_{s}$ ($q_{s}>0$). Starting
from the minimum possible value for $q_{v}/q_{s}$ (that one consonant with (%
\ref{22-3}) and (\ref{22a})), when only antiparticles levels show their
face, one sees that those levels tend to sink at the continuum of negative
energies while the particle levels emerge from the continuum of positive
energies. When $q_{v}/q_{s}=0$, the levels corresponding to particles and
antiparticles are disposed symmetrically about $E=0$. Another noteworthy
point is that for $q_{v}/q_{s}\geq 1$ ($q_{v}/q_{s}\leq 1$) there are only
particle (antiparticle) levels. On the other side, Figs. \ref{Fig2} and \ref
{Fig3} show the eigenenergies as a function of $q_{s}/q_{v}$. In both of
these last figures one can perceive the tendency of the spectra to be
symmetric about $E=0$ as $q_{s}/q_{v}\rightarrow \infty $. In particular,
notice in Fig. \ref{Fig2} that there are only particle levels in the event
that $-1<q_{s}/q_{v}<+1$.

\begin{figure}[th]
\begin{center}
\includegraphics[width=9cm, angle=270]{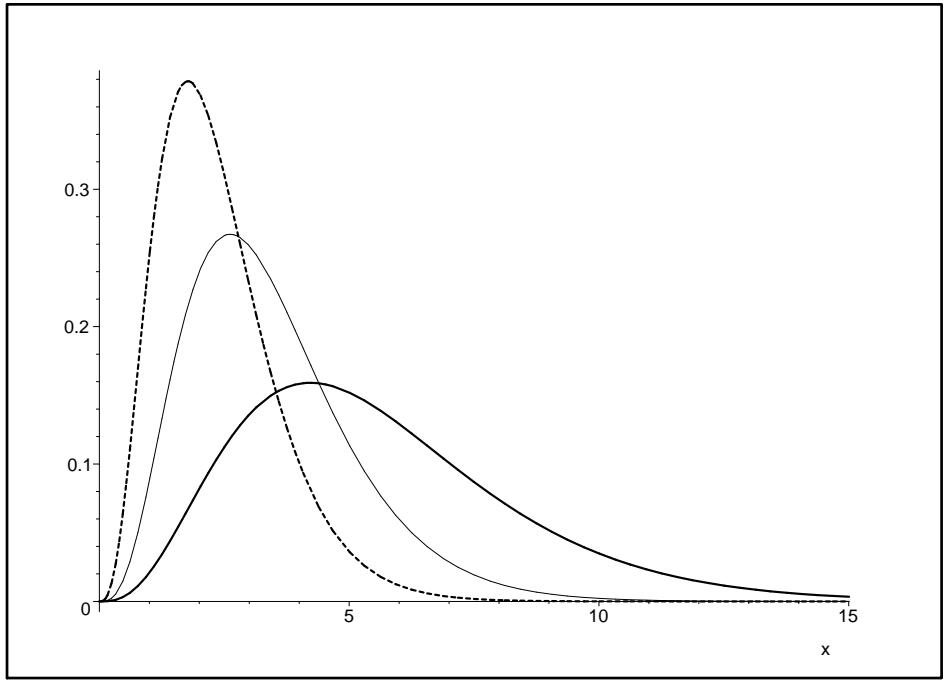}
\end{center}
\par
\vspace*{-0.1cm}
\caption{$|\psi |^{2}$ as a function of $x$, corresponding to the \textit{%
positive}-ground-state energy ($n=1$). The full thick line stands for $q_{v}
/q_{s}=-1/2$, the full thin line for $q_{v} /q_{s}=0$ and the dashed line
for $q_{v} /q_{s}=+1/2$ ($m=c=\hbar =q_{s}=1$).}
\label{Fig4}
\end{figure}

\begin{figure}[th]
\begin{center}
\includegraphics[width=9cm, angle=270]{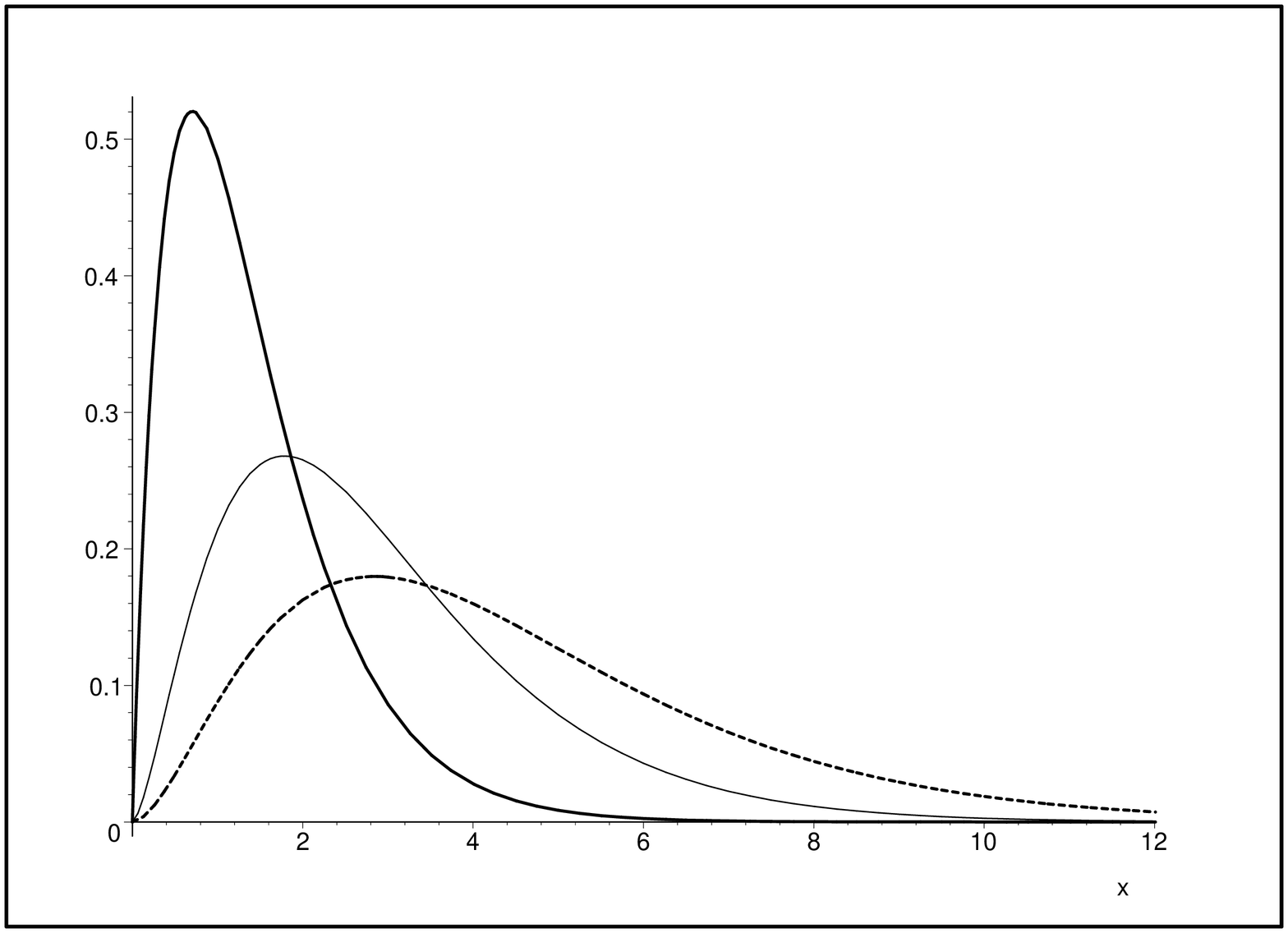}
\end{center}
\par
\vspace*{-0.1cm}
\caption{The same as in Fig. 4, but $q_{s}=0$. The full thick line stands
for $q_{v} =0.5$, the full thin line for $q_{v} =0.4$ and the dashed line
for $q_{v}=0.3$.}
\label{Fig5}
\end{figure}

Figs. \ref{Fig4} and \ref{Fig5} illustrate the behavior of the position
probability density associated to the Klein-Gordon ground-state
eigenfunction, $|\psi |^{2}$, on the positive side of the $x$-axis for the
positive-energy solutions. The normalization was obtained by numerical
computation. In Fig. \ref{Fig4} one can note that the position probability
density for the particle (antiparticle) states is more concentrated near the
origin for $q_{v}>0$ ($q_{v}<0$). In the limit $q_{v}\rightarrow q_{s}$ ($%
q_{v}\rightarrow -q_{s}$) the position probability densities corresponding
to the antiparticles (particles) spread without bound and only those ones
corresponding to the particles (antiparticles) are left. Comparison of the
different curves in Fig. \ref{Fig5} (the case of a pure vector coupling)
shows that the position probability density is notably smaller for smaller
values of $q_{v}$, and that the best localization of the particle is reached
for the highest possible value for the coupling constant.

Since the inversely linear potential given by (\ref{12}) is invariant under
reflection through the origin ($x\rightarrow -x$), eigenfunctions of the
wave equation given by (\ref{1b}) with well-defined parities can be found.
Those eigenfunctions can be constructed by taking symmetric and
antisymmetric linear combinations of $\psi $. These new eigenfunctions
possess the same Klein-Gordon eigenvalue, then there is a two-fold
degeneracy. Nevertheless, the matter is a little more complicated because
the effective potential presents a singularity. Recall that $\psi $ vanishes
at the origin but its first derivative does not, so the symmetric
combination of $\psi $ presents a discontinuous first derivative at the
origin. \noindent In fact, the second-order differential equation given by (%
\ref{1b}) implies that $\psi ^{\prime }$ can be discontinuous wherever the
potential undergoes an infinite jump. In the specific case under
consideration, the effect of the singularity of the potential can be
evaluated by integrating (\ref{1b}) from $-\delta $ to $+\delta $ and taking
the limit $\delta \rightarrow 0$. The connection condition among $\psi
^{\prime }(+\delta )$ and $\psi ^{\prime }(-\delta )$ can be summarized as

\begin{equation}
\psi ^{\prime }(+\delta )-\psi ^{\prime }(-\delta )=-\frac{cq_{eff}}{2\hbar }%
\int_{-\delta }^{+\delta }dx\;\frac{\psi }{|x|}  \label{23f}
\end{equation}
Substitution of (\ref{22}) into (\ref{23f}) allows us to conclude that $\psi
^{\prime }(+\delta )=\psi ^{\prime }(-\delta )$ in all the circumstances so
that we are forced to conclude that the Klein-Gordon eigenfunction must be
an odd-parity function. Therefore, the bound-state solutions are
nondegenerate.

We have succeed in searching for exact Klein-Gordon bounded solutions for
massive particles by considering a mixing of vector-scalar inversely linear
potentials in 1+1 dimensions. For $q_{s}^{2}>q_{v}^{2}$, there exist
bound-state solutions for particles and antiparticles. As for $q_{s}^{2}\leq
q_{v}^{2}$, there are bound-state solutions either for particles or for
antiparticles. Invariably, the spectrum is nondegenerate and the
eigenfunction behaves as an odd-parity function. In addition, for the
special case $q_{s}>-q_{v}$ the theory presents a definite nonrelativistic
limit ($|q_{s}|,|q_{v}|\ll 1$ and $E\simeq mc^{2}$), as one should expect.

Beyond its intrinsic importance as a new solution for a fundamental equation
in physics, the problem analyzed in this paper favors the conclusion that
even-parity solutions to the nonrelativistic one-dimensional hydrogen atom
do not exist.

\bigskip \bigskip \bigskip \bigskip

\vspace{1in}

\noindent{\textbf{Acknowledgments} }

This work was supported in part by means of funds provided by CNPq and
FAPESP.

\newpage

\newpage

\begin{thebibliography}{99}
\bibitem{xia}  D. Xianxi et al., Phys. Rev. A 55 (1997) 2617 and references
therein.

\bibitem{lou}  R. Loudon, Am. J. Phys. 27 (1959) 649.

\bibitem{spe}  H.N. Spector, J. Lee, Am. J. Phys. 53 (1985) 248.

\bibitem{mos}  R.E. Moss, Am. J. Phys. 55 (1987) 397.

\bibitem{ho}  C.-L. Ho, V.R. Khalilov, Phys. Rev. D 63 (2001) 027701.

\bibitem{asc6}  A.S. de Castro, Phys. Lett. A 328 (2004) 289.

\bibitem{kra}  A. Kratzer, Z. Phys. 3 (1920) 289.

\bibitem{coo}  F. Cooper et al., Ann. Phys. (N.Y.) 187 (1988) 1.

\bibitem{nog}  Y. Nogami, F.M. Toyama, Phys. Rev. A 47 (1993) 1708.

\bibitem{col1}  S. Coleman et al., Ann. Phys. (N.Y.) 93 (1975) 267.

\bibitem{col2}  S. Coleman, Ann. Phys. (N.Y.) 101 (1976) 239.

\bibitem{fro}  J. Fr\"{o}hlich, E. Seiler, Helv. Phys. Acta 49 (1976) 889.

\bibitem{cap}  A.Z. Capri, R. Ferrari, Can. J. Phys. 63 (1985) 1029.

\bibitem{gal}  H. Gali\'{c}, Am. J. Phys. 56 (1988) 312.

\bibitem{hoo}  G.\'{}t Hooft, Nucl. Phys. B 75 (1974) 461$.$

\bibitem{kog}  J. Kogut, L. Susskind, Phys. Rev. D 9 (1974) 3501.

\bibitem{bha}  R.S. Bhalerao, B. Ram, Am. J. Phys. 69 (2001) 817.

\bibitem{cas1}  A.S. de Castro, Am. J. Phys. 70 (2002) 450.

\bibitem{cav}  R.M. Cavalcanti, Am. J. Phys. 70(2002) 451.

\bibitem{hil}  J.R. Hiller, Am. J. Phys. 70 (2002) 522.

\bibitem{luc}  W. Lucha et al., Phys. Rep. 200 (1991) 127 and references
therein.

\bibitem{ser}  B.D. Serot, J.D. Walecka, in: Advances in Nuclear Physics,
Vol. 16, edited by J.W. Negele and E. Vogt, Plenum, New York, 1986.

\bibitem{gin3}  J.N. Ginocchio, Phys. Rev. Lett. 78 (1997) 436.

\bibitem{gin}  J.N. Ginocchio, A. Leviatan, Phys. Lett. B 425 (1998) 1.

\bibitem{gin4}  J.N. Ginocchio, Phys. Rep. 315 (1999) 231.

\bibitem{alb1}  P. Alberto et al., Phys. Rev. Lett. 86 (2001) 5015.

\bibitem{alb2}  P. Alberto et al., Phys. Rev. C 65 (2002) 034307.

\bibitem{che}  T.-S. Chen et al., Chin. Phys. Lett. 20 (2003) 358.

\bibitem{mao}  G. Mao, Phys. Rev. C 67 (2003) 044318.

\bibitem{nosso}  R. Lisboa et al., Phys. Rev. C 69 (2004) 024319.

\bibitem{cas2}  A.S. de Castro, Phys. Lett. A 305 (2002) 100.

\bibitem{ntd}  Y. Nogami et al., Am. J. Phys. 71 (2003) 950.

\bibitem{asc7}  A.S. de Castro, to be published in Ann. Phys. (N.Y.).

\bibitem{lan}  L.D. Landau, E.M. Lifshitz, Quantum Mechanics, Pergamon, N.
York, 1958.

\bibitem{bag}  V.G. Bagrov, D.M. Gitman, Exact Solutions of Relativistic
Wave Equations, Kluer, Dordrecht, 1990.

\bibitem{flu}  S. Fl\"{u}gge, Practical Quantum Mechanics, Springer-Verlag,
Berlin, 1999.

\bibitem{abr}  M. Abramowitz, I.A. Stegun, Handbook of Mathematical
Functions, Dover, Toronto, 1965.
\end{thebibliography}
\end{document}